\documentclass[a4paper]{jpconf}
\usepackage{graphicx}
\usepackage{graphics}
\usepackage{times}
\usepackage{amsmath}
\usepackage{citesort}
\usepackage{comment} 

\def\agt{\mathrel{\raise.3ex\hbox{$>$}\mkern-14mu\lower0.6ex\hbox{$\sim$}}}
\def\alt{\mathrel{\raise.3ex\hbox{$<$}\mkern-14mu\lower0.6ex\hbox{$\sim$}}}
\def\prd{\textit{Phys. Rev. }D}
\def\apj{\textit{Ap. J.}}
\def\prl{\textit{Phys. Rev. Lett.}}
\usepackage[normalem]{ulem}
\usepackage{amssymb}

\begin{document}

\title{Neutron star equation of state via gravitational wave observations}

\author{C Markakis$^1$, J S Read$^2$, M Shibata$^3$, K Ury\=u$^4$, J D E Creighton$^1$, J L Friedman$^1$, and B D Lackey$^1$}
\address{$^1$ Department of Physics, University of Wisconsin--Milwaukee, P.O. Box 413, Milwaukee, WI 53201, USA}
\address{$^2$ Max-Planck-Institut f\"{u}r Gravitationsphysik, Albert-Einstein-Institut, Golm, Germany}
\address{$^3$ Yukawa Institute for Theoretical Physics,
Kyoto University, Kyoto 606-8502, Japan}
\address{$^4$ Department of Physics, University of the Ryukyus, 1 Senbaru, Nishihara, Okinawa 903-0213, Japan}
%\date[\relax]{RCS \thercsid; compiled \today}
%\pacs{04.25.dk, 04.30.Tv, 04.25.Nx, 26.60.Kp, 04.80.Nn}
\begin{abstract}
Gravitational wave observations can potentially measure properties of neutron star equations of state by measuring departures from the point-particle limit of the gravitational waveform produced in the late inspiral of a neutron star binary. Numerical simulations of inspiraling neutron star binaries computed for equations of state with varying stiffness are compared. As the stars approach their final plunge and merger, the gravitational wave phase accumulates more rapidly if the neutron stars are more compact. This suggests that gravitational wave observations at frequencies around 1 kHz will be able to measure a compactness parameter and place stringent bounds on possible neutron star equations of state. Advanced laser interferometric gravitational wave observatories will be able to tune their frequency band to optimize sensitivity in the required frequency range to make sensitive measures of the late-inspiral phase of the coalescence. 

% Because these measurements make use of the finite-size effects on the late inspiral
% rather than the oscillations of a quasi-stable final neutron star remnant,
% careful numerical treatments of temperature, neutrino transport, differential rotation,
% magnetic fields, are not important to the qualitative results
% presented here.  
\end{abstract} %\maketitle

\section{Introduction}

In numerical simulations of the late inspiral and merger of 
binary neutron-star systems, one commonly specifies an equation 
of state  for the matter, perform a numerical simulation and extract the gravitational waveforms produced in the inspiral.
The scope of this talk is to report work on the inverse problem: if  gravitational radiation from  an inspiraling binary system is detected, can one use it to infer the bulk properties of neutron star matter and, if so, with what accuracy?
To answer this question,
a number of simulations with systematic variation of the equation of state (EOS) is performed.
The simulations use the evolution and initial-data
codes of Shibata and Ury\=u to compute the last several orbits and the
merger of neutron stars, with matter described by a parameterized
equation of state, previously obtained in \cite{pppp}.
Our analysis uses  waveforms  from these simulations  to examine the accuracy with which detectors with the sensitivity of
Advanced LIGO can extract from inspiral waveforms an EOS
parameter associated with the stiffness of the neutron star EOS
above nuclear density. 

 One might expect that  gravitational wave observations can potentially measure properties of
neutron star EOS by measuring departures from the
point-particle limit of the gravitational waveform produced
during the late inspiral of a binary system. A  way to quantify this departure is to make use of the fact that the gravitational wave phase near the end of inspiral accumulates more rapidly
for smaller values of the neutron star compactness (the ratio of
the mass of a neutron star to its radius). In this way, greater accuracy may be obtained compared to  previous work that suggested the use of an effective cutoff
frequency to place constraints on the equation of state. 
%%The signal analysis focuses on the late inspiral, as the
%%radius of the orbit $r$ approaches the neutron star radius $R$. The orbital
%%dynamics in this region will depend on the radius and internal structure
%%of the neutron star, which in turn depend on the EOS\@.
 Our results based on this approach suggest that realistic equations of state will lead to gravitational
waveforms that are distinguishable from point particle inspirals at a distance  $\lesssim 100\,\rm{Mpc}$ for an optimally oriented system, using Advanced LIGO in a broadband or narrowband  configuration. Waveform analysis, that uses the sensitivity curves of commissioned and
proposed gravitational wave detectors, allows us to also estimate the
accuracy with which neutron star radii, closely linked to the EOS parameter
varied, can be extracted.
The choice of EOS parameter varied in this work is motivated by the fact that, as Lattimer and
Prakash observed \cite{lattprak}, neutron-star radius is closely tied to the pressure
at density not far above nuclear. Analysis of the numerical waveforms also indicates that optimizing the  sensitivity of gravitational wave detectors  to frequencies above 700 Hz can lead to improved constraints on the radius and EOS\    of neutron stars.

\section{Equation of state} \label{sec:modeleos}

 Difficulties inherent in astrophysically constraining the EOS of neutron star matter   are the wide range of different microphysical models, implying the lack of a model-independent set of fundamental observables, and  the fact that models often have more free parameters than the number of astrophysical observables. Therefore, an effort to systematize such constraints necessitates
a form of EOS (i) parameterized in a model independent way,
with a number of EOS parameters (ii) large enough to  accurately model the universe of candidate EOS, but (iii) smaller than the number of 
neutron star observables that have been measured or will have been measured in the next several years.

In the past, for different purposes,  M\"uller and
Eriguchi \cite{MullerEriguchi} have used non-relativistic piecewise polytropes, with a large number (20-100) of segments, to accurately approximate realistic  EOS,    in order to  construct Newtonian models of differentially rotating  stars. More recently, with conditions (i), (ii) and especially (iii) in mind, Read et al  \cite{pppp,ReadThesis2008} have developed a relativistic piecewise polytropic approximation, which  reproduces the features of most realistic neutron star EOS with rms error $\lesssim4\%$, using only 3-4 segments (or 3-4 free parameters),  as described below.

\subsection{  Astrophysical constraints on a piecewise polytropic equation of state}

The EOS pressure $p$ is specified
as a function of rest mass density $\rho$
(rest mass density $\rho = m_{\text{b}} n$ is proportional to
number density $n$, where $m_{\text{b}} = 1.66\times
10^{-24}$\,g is the mean baryon rest mass). The
relativistic piecewise polytropic EOS  proposed in \cite{pppp,ReadThesis2008} has the form \begin{equation} \label{eq:eos}
p(\rho) = K_{i} \rho^{\Gamma_{i}}
\end{equation} 
in a set of intervals $\rho_{i-1} \leq \rho \leq \rho_{i}$ in rest mass
density, with the constants $K_i$  determined by requiring  continuity on each dividing density $\rho_i$:
\begin{equation}
K_{i}\rho_i^{\Gamma_{i}}=K_{i+1} \rho_i^{\Gamma_{i+1}}.
\end{equation}
The  energy density $\epsilon$ as a function of $\rho$ is then
determined by the first law of thermodynamics,
\begin{equation} \label{eq:eos3}
d\frac\epsilon\rho = -p \,
d\frac1\rho, 
\end{equation}
which is integrated, to give
\begin{equation} \label{eq:eos3}
\epsilon(\rho)=(1+a_i)\rho c^2+\frac{K_i \rho^{\Gamma_{i}}}{\Gamma_i-1} 
\end{equation}
(for $\Gamma_i\neq1$), where we used  the condition $\epsilon / \rho \to c^2$ as $\rho \to 0$. This condition implies $a_0=0$ and the other  constants $a_i$ are fixed by continuity of   $\epsilon(\rho)$ at each  $\rho_i$. In the above equations, $\rho$
is measured in g\,cm$^{-3}$, $\epsilon$ and $p$ have units of erg\,cm$^{-3}$\,$=$\,dyn\,cm$^{-2}$
and
%$c=2.99792458\times10^{10}$\,cm/s
$c$
is the speed of light. The specific enthalpy, $\mathfrak{h}$, is dimensionless and is given by
\begin{equation} \label{eq:specificenthalpy}
\mathfrak{h}\ =\frac{\epsilon + p}{\rho c^2}
\end{equation}
The crust EOS is considered to be known and, for the purposes of this  work, can be modelled with a single polytropic  region,
fitted to a tabulated crust EOS for the region above neutron drip
($10^{11}$-$10^{12}$\,g\,cm$^{-3}$); the numerical simulations considered do not resolve
densities below this. This
polytrope has $\Gamma_{0}\equiv\Gamma_{\text{crust}} = 1.3569$, with $K_0\equiv K_{\text{crust}}$
chosen so that $p= 1.5689\times10^{31}$\,dyn\,cm$^{-2}$ when
$\rho=10^{13}$\,g\,cm$^{-3}$. The core EOS is constructed independently
of crust behavior.  In \cite{pppp} it was found that three zones within the core, with adiabatic exponents $\Gamma_1, \Gamma_2, \Gamma_3 $ as  parameters, are
needed to  model the full range of proposed EOS models with   sufficient accuracy.  A fourth parameter, $p_1\equiv p(\rho_1)$, is needed to fix an overall pressure shift, specified at the fiducial density $\rho_1$.
The dividing densities $\rho_1,\rho_2$  are \textit{not} considered new
parameters.
Instead, they are fixed by minimizing the error in fitting a large set of
candidate EOS, leading  \cite{pppp} to the preferred values
$\rho_1 = 10^{14.7}$\,g\,cm$^{-3}$ 
and  $\rho_2 =
10^{15.0}$\,g\,cm$^{-3}$. 
The first dividing density $\rho_0$ between the crust and core
varies by  EOS, but is also not a new parameter. It is defined as the density for which the $\log P$ vs. $\log \rho$ curves of the crust and core EOS intersect.
 Fixing the crust EOS and determining the three dividing densities in this way, allows one to model a large set of candidate EOS with only four
parameters $\{p_1, \Gamma_1, \Gamma_2, \Gamma_3\}$.
%(this excludes EOSs with values of $p_1$ and $\Gamma_1$ that are
%incompatible, for which the slope of the  $\log P$ vs. $\log \rho$ 
%curve is too shallow to reach the pressure $p_1$ from the
%low-density part of the EOS)

Using this parameterized EOS, one can construct relativistic models
of neutron stars and compute astrophysical observables, such as maximum mass,  radius, moment of inertia etc. Doing so, while systematically varying the EOS parameters $\{p_1, \Gamma_1, \Gamma_2, \Gamma_3\}$, Read et al \cite{pppp}
have obtained constraints on the allowed parameter
space set by causality [Fig.~1] and by present and near-future astronomical observation.
 They find that two
measured properties of a single star
can potentially place  stringent constraints on the EOS parameter space  and that a potential measurement of the moment of inertia of
a pulsar can strongly constrain the maximum neutron-star mass [Figs.~1,~2].
A detailed study of  these constraints may be found in  \cite{pppp,ReadThesis2008}.
  The potential combination of such astrophysical constraints with  gravitational wave observation can restrict the allowed EOS parameter space region to a
surface (thickened by the error bars of 
the observations) or even a line. The rest of this talk focuses on constraints  placed on neutron star EOS by potential  detection of binary inspiral waveforms.

\begin{figure}[!htb]
\begin{center}
\begin{tabular}{cc}
\includegraphics[height=50mm]{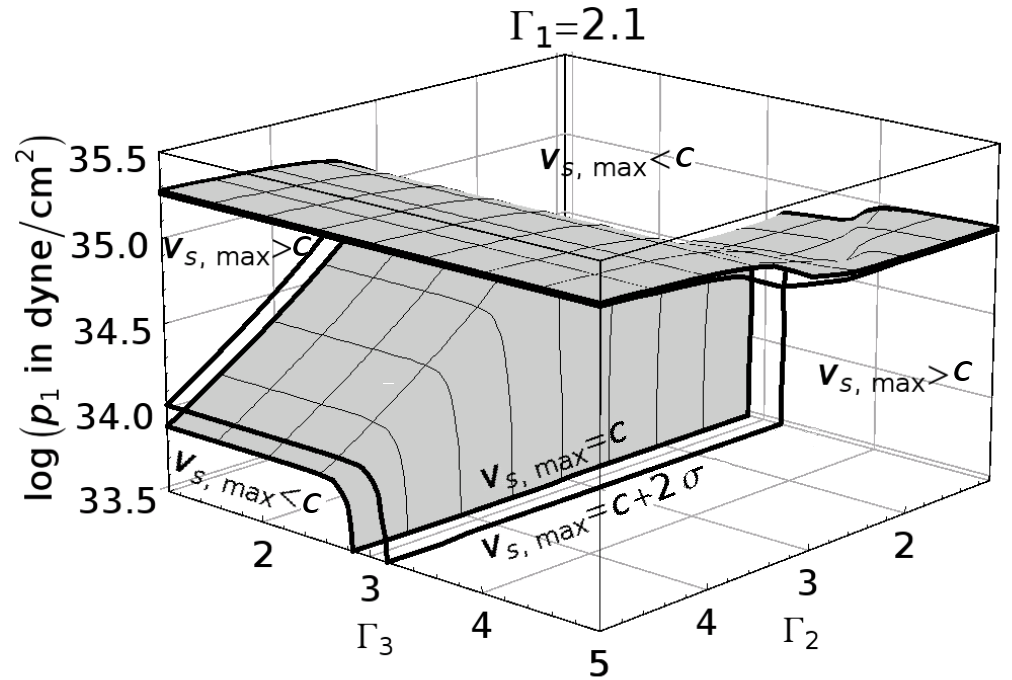}
\includegraphics[height=47mm]{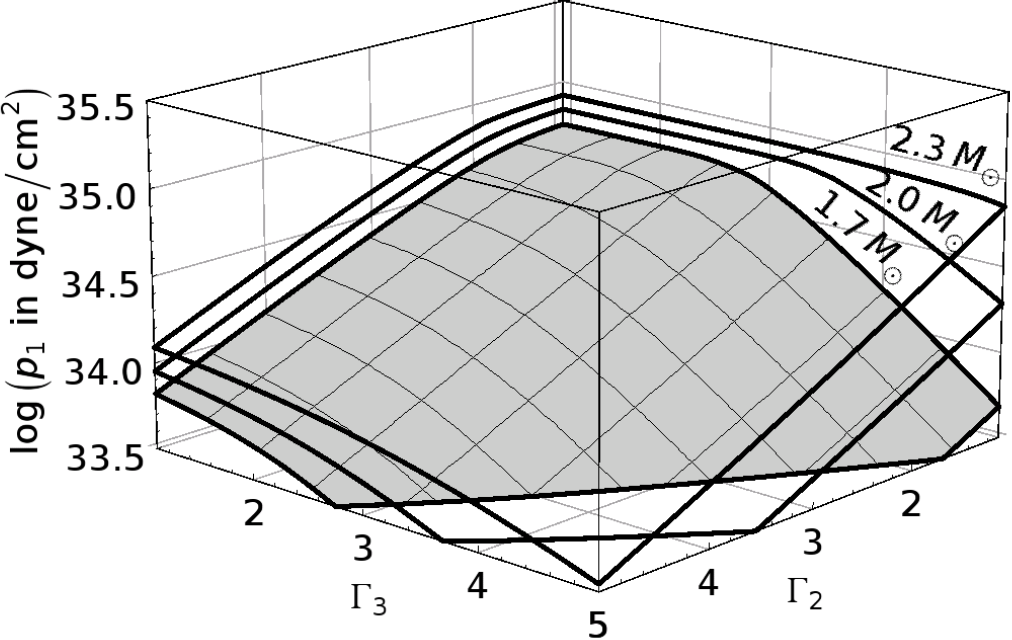} 
\end{tabular}
\end{center}
\caption[EOS for numerical evolution]{
\label{fig:b1} \textit{Left}: 
Causality constraints are shown for a fixed value of $\Gamma_1$. For each EOS, the maximum speed of sound up to the central density of the maximum mass star is considered. The shaded surface separates the EOS parameter space into a region \textit{behind} the surface allowed by causality, labeled $v_s<c$, and  a region in which corresponding EOSs violate causality, labeled $v_s>c$. The second, outlined surface shows a weaker constraint to accommodate the expected error ($\sim2$ standard deviations $\sigma$) 
in the speed of sound associated with a piecewise polytropic approximation to an EOS.\\
\textit{Right}: Surfaces representing the set of parameters  resulting in a constant maximum mass. A single observation of a massive neutron star constrains the equation of state to lie \textit{above} the corresponding surface. The exponent $\Gamma_1$ is set to the least constraining value at each point (figures adopted from \cite{pppp}).}
\end{figure}

\begin{figure}[!htb]
\begin{center}
\begin{tabular}{cc}
\includegraphics[height=46mm]{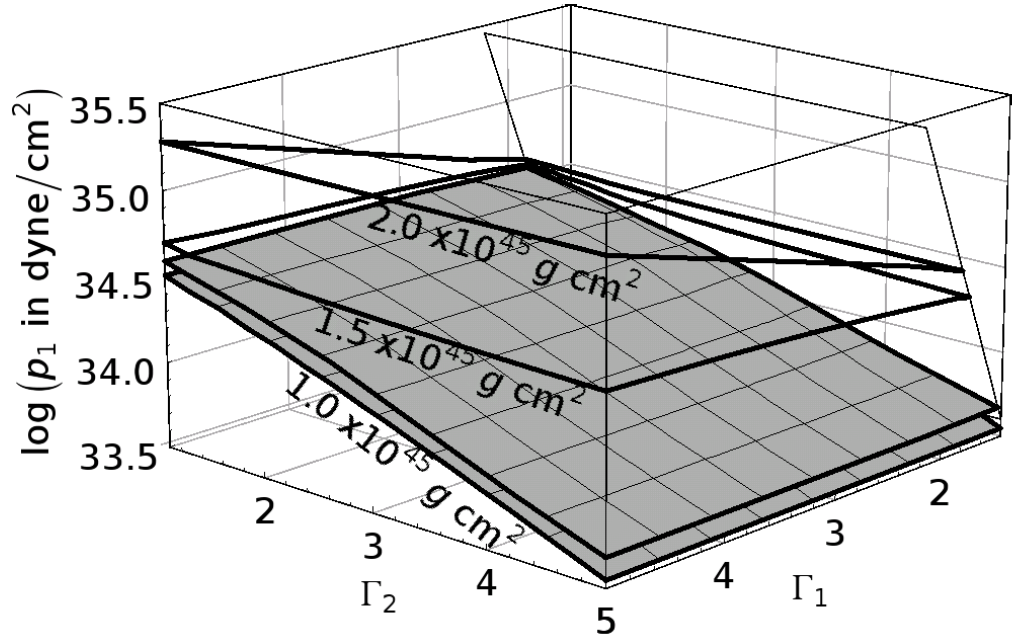} 
\includegraphics[height=45mm]{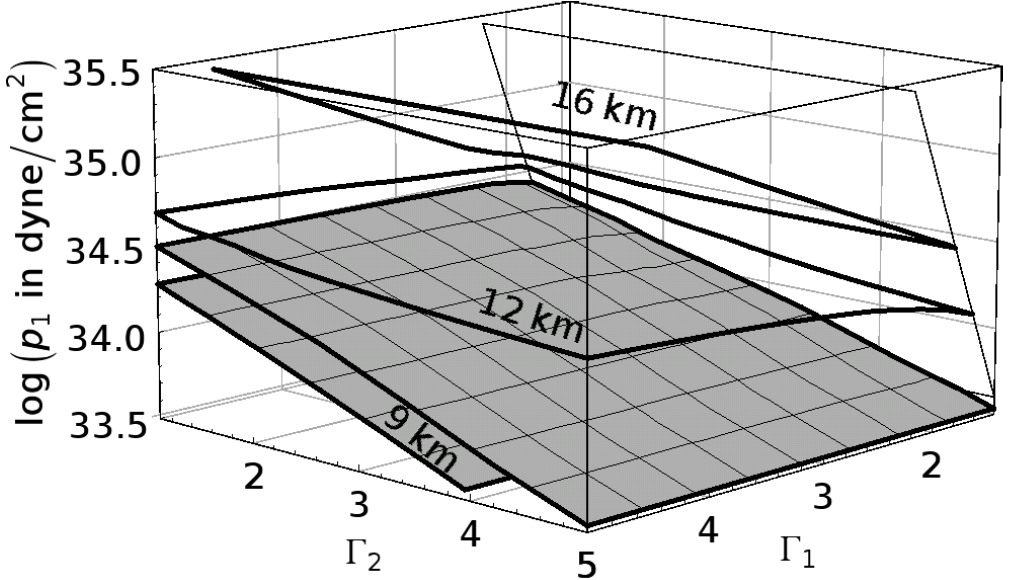}
\end{tabular}
\end{center}
\caption[EOS for numerical evolution]{
\label{fig:b23} \textit{Left}: 
  Surfaces that represent the set of parameters
that result in a star with a mass of 1.338\,$M_\odot$ and a given moment of inertia $I$ (from possible near-future observations of PSR J0737-3039A).
 Observation of moment of inertia $I$ for a known mass $M$ constrains the equation of state to lie \textit{on} the corresponding surface.
\\
\textit{Right}:
Surfaces representing the set of parameters
that result in a star with a mass of 1.4\,$M_\odot$ and a fixed
radius $R$. Measurement of the radius $R$ of a neutron star with known mass $M$ constrains the equation of state to lie \textit{on} the corresponding surface. The wedge at the back right in both figures corresponds
to incompatible values of $p_1$ and $\Gamma_1$ (figures adopted from \cite{pppp}).}
\end{figure}

%%\begin{figure}[!htb]
%%\begin{center}
%\includegraphics[height=50mm]{02mall.png} 
%%\begin{tabular}{cc}
%%\includegraphics[height=46mm]{03i1338.png} 
%\includegraphics[height=47mm]{02mall.png} 
%%\includegraphics[height=45mm]{05r14.png} 
%%\\
%%\end{tabular}

%%\end{center}
%%\caption[EOS for numerical evolution]{
%%\label{fig:b23}\textbf{} 
%%\textit{Left}: Surfaces ....
%%\\\textit{
%%Right}:
%% ... (figures adopted from \cite{pppp}).
%%}
%%\end{figure}

%%\begin{figure}[!htb]
%%\begin{center}

%%\begin{tabular}{cc}

%%\includegraphics[height=47mm]{04imvs3d.png} 
%%&
%%\includegraphics[height=47mm]{04imvs3d2.png} 
%%\\
%%\end{tabular}
%%\end{center}

%%\caption[EOS for numerical evolution]{
%%\label{fig:b5}\textbf{}??? Surfaces ... [might delete this]
%%(figure adopted from \cite{pppp}).
%%}
%%\end{figure}

%%\begin{figure}[!htb]
%%\begin{center}
%%\includegraphics[height=50mm]{05r14.png} 
%%\end{center}
%%\caption[EOS for numerical evolution]{
%%\label{fig:b6}\textbf{} ... (figure adopted from \cite{pppp}).
%%}
%%\end{figure}

\subsection{Initial candidate equations of state}

Systematic variation of the EOS parameters
also allows us to determine which properties significantly affect the gravitational radiation
produced, and thus can be constrained with detection of sufficiently strong
gravitational waves.

The models selected for this study use a variation of one EOS parameter in the
neutron star core. 
We vary the core EOS by an overall pressure shift $p_1$
(specified at the fiducial density
$\rho_1 = 10^{14.7}$\,g\,cm$^{-3}$), while holding the adiabatic index in
the core regions fixed at $\Gamma_1=\Gamma_2=\Gamma_3=3$.  
While only a subset of realistic EOS are well-approximated by a single core
polytrope, reducing the EOS considered to this single-parameter family
allows us to estimate parameter measurability with a reasonable
number of simulations.

\begin{figure}[!htb]
\begin{center}
\includegraphics[height=80mm]{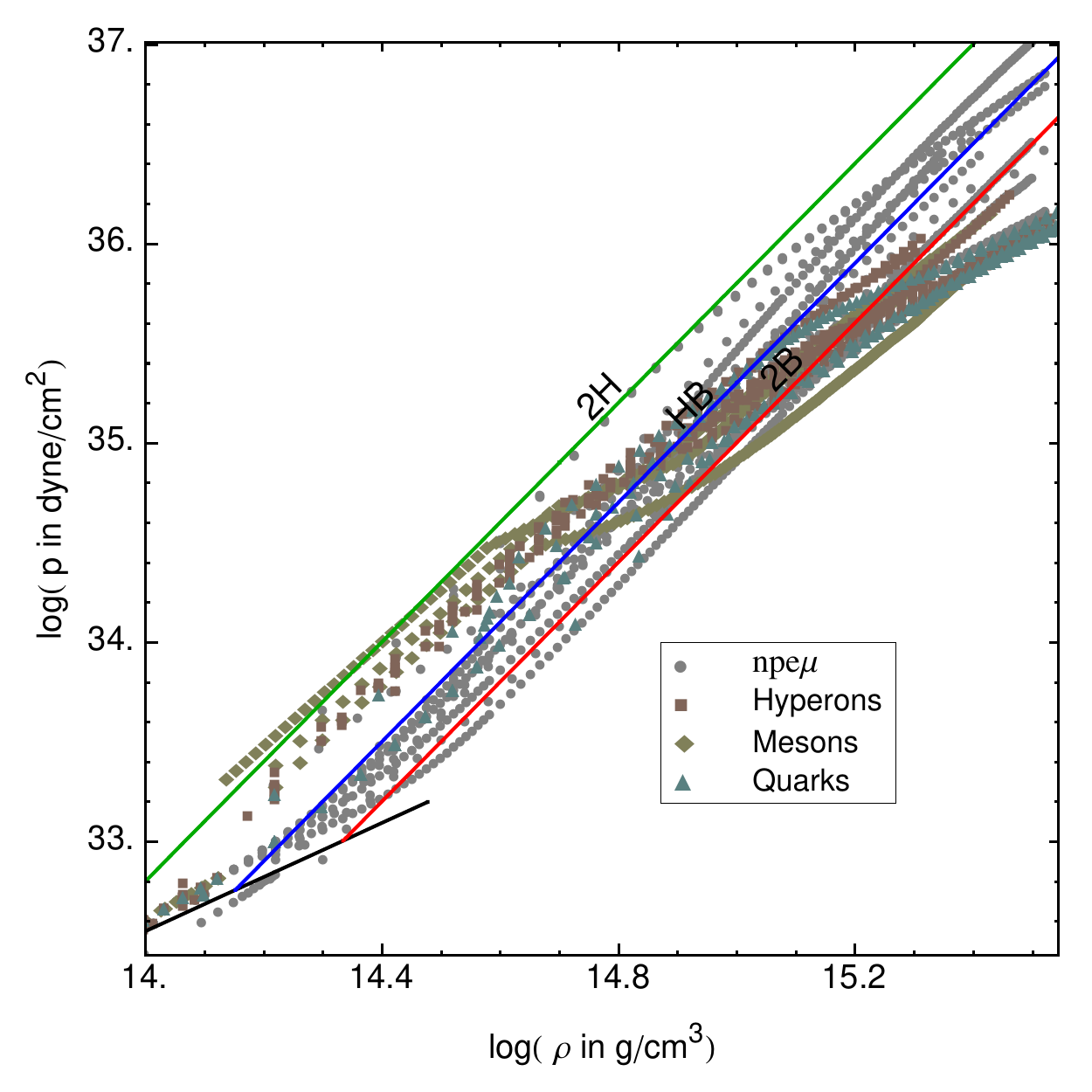} 
\end{center}
\caption[EOS for numerical evolution]{
\label{fig:candidates}\textbf{} Initial choices of EOS for numerical evolution 
compared to the set of tabled EOS considered in \cite{pppp}.
Candidates are labelled in order of increasing softness: 2H,
% H,
  HB,
%   B,
    2B. Reproduced with permission from  \cite{Readetal2009} \copyright (2009) by the American Physical Society.}
\end{figure}

After fixing the core adiabatic index, increasing the overall pressure
scale $p_1$ produces a family of neutron stars with progressively
increasing radius for a given mass; the choice of parameter $p_1$ is partly motivated by the observation of Lattimer and Prakash \cite{lattprak} that pressure at one to
two times nuclear density is closely tied to neutron star radius, with $R
\propto p_1^{1/4}$.  The radius is less sensitive to variation of the
adiabatic index in the neutron star core, for reasonable adiabatic indices
\cite{pppp}, as suggested by Fig.~\ref{fig:b23}.
 
Future work will incorporate additional
variations of the EOS within the core. This could involve additional models of
 varying adiabatic index around a fixed $p_1$  (with $\Gamma_1=\Gamma_2=\Gamma_3$ in all core layers), as well as multiple
piecewise-polytrope zones 
(with $\Gamma_1\neq\Gamma_2\neq\Gamma_3$) 
within the core or EOS parameters yielding
neutron stars of the same mass $M$ and radius $R$ but different internal structure.  Such work
would yield insight into the relative size and correlation of effects on
the orbital evolution due to the stellar radius \textit{and} internal
structure.

The first set of EOS were chosen to ``bracket'' the range of
existing candidates, seen in Fig.~\ref{fig:candidates}. These models
are HB with $p_1 =10^{34.40}\,\text{dyn}\,\text{cm}^{-2}$, a standard
EOS; 2H with $p_1=10^{34.90}\,\text{dyn}\,\text{cm}^{-2}$, a stiff
EOS; 2B, with $p_1=10^{34.10}\,\text{dyn}\,\text{cm}^{-2}$, a soft
EOS\@. 
%Additional models B with
%$p_1/c^2=10^{34.30}\,\text{dyn}\,\text{cm}^{-2}$ and H with
%$p_1/c^2=10^{34.50}\,\text{dyn}\,\text{cm}^{-2}$, were chosen with
%small shifts in parameter from HB to better estimate local parameter
%dependence of the waveform.

%Although the set of models is explicitly constructed by varying the single
%parameter $p_1$,  the resulting neutron star models also have systematic
%variation in the radius $R$, as seen in Table \ref{tab:modprop}. They can
%also be considered a one-parameter family of varying $R$; in fact, $p_1$
%was chosen as an EOS parameter as Lattimer and . The
%parameter estimation analysis is done for both $p_1$ and $R$. 

%We expect the structure of $\sim 1.4 M_\odot$ neutron
%stars to depend most strongly on the value of $p_1$, based on both the
%empirical formula of Lattimer and Prakash \cite{lattprak} describing
%radius as a function of pressure at a fiducial density at some $\rho \simeq
%2.7$ to $5 \times10^14$\,g\,cm$^{-3}$, and the piecewise polytrope equation
%of state study described in Chapter \ref{ch:ppp} showing moment of inertia
%of $1.388 M_\odot$ neutron stars is largely dependent on $p_1$.

\section{Numerical simulations}

For each parameterized EOS considered, the late inspiral and
merger of a binary neutron star system is simulated. The
gravitational mass of each neutron star in the binary is fixed to $1.35 M_\odot$, an
average value for pulsars observed in binary systems
\cite{ThorsettChakrabarty,DoublePulsar2}. The significance of
tidal effects in this configuration is expected to fairly represent tidal
effects over the relatively narrow range of masses and mass ratios  in
astrophysical binary neutron star systems.

\subsection{Initial data}

Initial data is generated by constructing quasi-equilibrium sequences of irrotational neutron stars in a binary system,
following
the methods of \cite{BNSinitial1,BNSinitial5,BNSinitial,BNSinitial2,BNSinitial3,BNSinitial4}, using the
Isenberg-Wilson-Mathews formulation \cite{IWM1,IWM2}. %As in previous work, we assume equal-mass binaries, and
 Since tidal spin-up in neutron star binaries is estimated to be negligible   \cite{irrotBNS,irrotBNS2}, our initial data and quasi-equilibrium sequences are constructed assuming irrotational flow fields and neglecting the neutron star spin.\ The parameterized
EOS of Eq.~\eqref{eq:eos} is incorporated in the code to
solve for initial data with a conformally flat spatial geometry coupled to the fluid
equations of neutron star matter.  The baryon number of each star is equal to that of a spherical isolated  star with gravitational mass
$M=1.35\,M_\odot$. The initial data for the full numerical  simulation is
taken from the quasi-equilibrium configuration at a separation such that approximately $3$ orbits remain until merger. Relevant quantities of the initial
configurations for each parameterized EOS are presented in
Tables~\ref{tab:modprop},~\ref{tab:initial}.

%\begin{table}[!htb]
\begin{table*}[!htb]
\caption[Properties of EOSs for numerical simulation]{Properties of
initial candidate EOS\@. These range from the ``softest''
EOS at the top, which results in a prompt collapse to
a black hole upon merger, to the ``stiffest'' at the
bottom.  Model HB is considered a typical EOS\@.  The pressure
$p_1$, which is the pressure at density
$\rho_1=5\times10^{14}\,\text{g}\,\text{cm}^{-3}$, determines the polytropic
EOS for the neutron star core; all candidates have $\Gamma=3$.
Radius $R$ and compactness $GM/c^2R$ are those of a single isolated spherical (TOV) star of gravitational mass $M=1.35\,M_\odot$, with radius measured in Schwarzschild coordinates.
The maximum neutron
star mass $M_{\text{max}}$  is given in the fifth column.
\label{tab:modprop}}
\begin{center}
\begin{tabular}{lcccc}\hline
Model & $\log_{10} p_1\,[\mbox{dyn}\,\mbox{cm}^{-2}]$ & $R\,[\mbox{km}]$
& $GM/c^2R$ & $M_{\text{max}}\,[M_\odot]$\\
\hline\hline
2H & 34.90 & 15.2 & 0.13 & 2.83\\
%H  & 34.50 & 12.3 & 0.16 & 2.25\\
HB & 34.40 & 11.6 & 0.17 & 2.12\\
%B  & 34.30 & 10.9 & 0.18 & 2.00\\
2B & 34.10 & \phantom{0}9.7 &0.21 & 1.78 \\
\hline
\end{tabular}
\end{center}
\end{table*}

\begin{table*}[!htb]
\caption{Quantities of initial data sets for
irrotational binary neutron stars.  
Each star has a baryon number $M_0$ equal to that of an isolated star with gravitational mass
$M=1.35\,M_\odot$. The ADM mass $M_\text{ADM}$ of the initial slice
includes the binding energy and  $J$ is the total angular momentum of the
initial slice. 
%The binary compactness $C_0$ is defined
%by $C_0=(\Omega M_{\text{ADM}} Gc^{-3})^{2/3}$.
\\
}
\label{tab:initial}
\begin{center}
\begin{tabular}{lccccc}
\hline
Model&$\rho_{\text{max}}\,[\mbox{g}\,\mbox{cm}^{-3}]$ &$M_0 \,[M_\odot]$
&$M_{\text{ADM}}
\,[M_\odot]$& $cJ / ( G M_{\text{ADM}}^2)$&$\Omega / 2\pi$ [Hz] %& $C_0$ 
\\
\hline\hline
2H &$3.73196\times10^{14}$ &1.45488 & 2.67262 & 0.993319 & 324.704 %& $8.96966\times10^{-2}$
\\
%H  &$7.02661\times10^{14}$ &1.48385 & 2.67080 & 0.989524 & 321.468 &$8.90593\times10^{-2}$\\
HB &$8.27673\times10^{14}$ &1.49273 &2.67290 &0.995361 & 309.928 %& $8.69582\times10^{-2}$
\\
%B &$9.77811\times10^{14}$ &1.50247 & 2.67290 & 0.992638 & 314.170&$8.77522\times10^{-2}$\\
2B &$1.38300\times10^{15}$ &1.52509 & 2.67229 & 0.987681 &321.170 %& $8.90375\times10^{-2}$
\\
\hline
\end{tabular}
\end{center}
\end{table*}

\begin{figure}[!htb]
\begin{center}
\includegraphics[width=.7\textwidth]{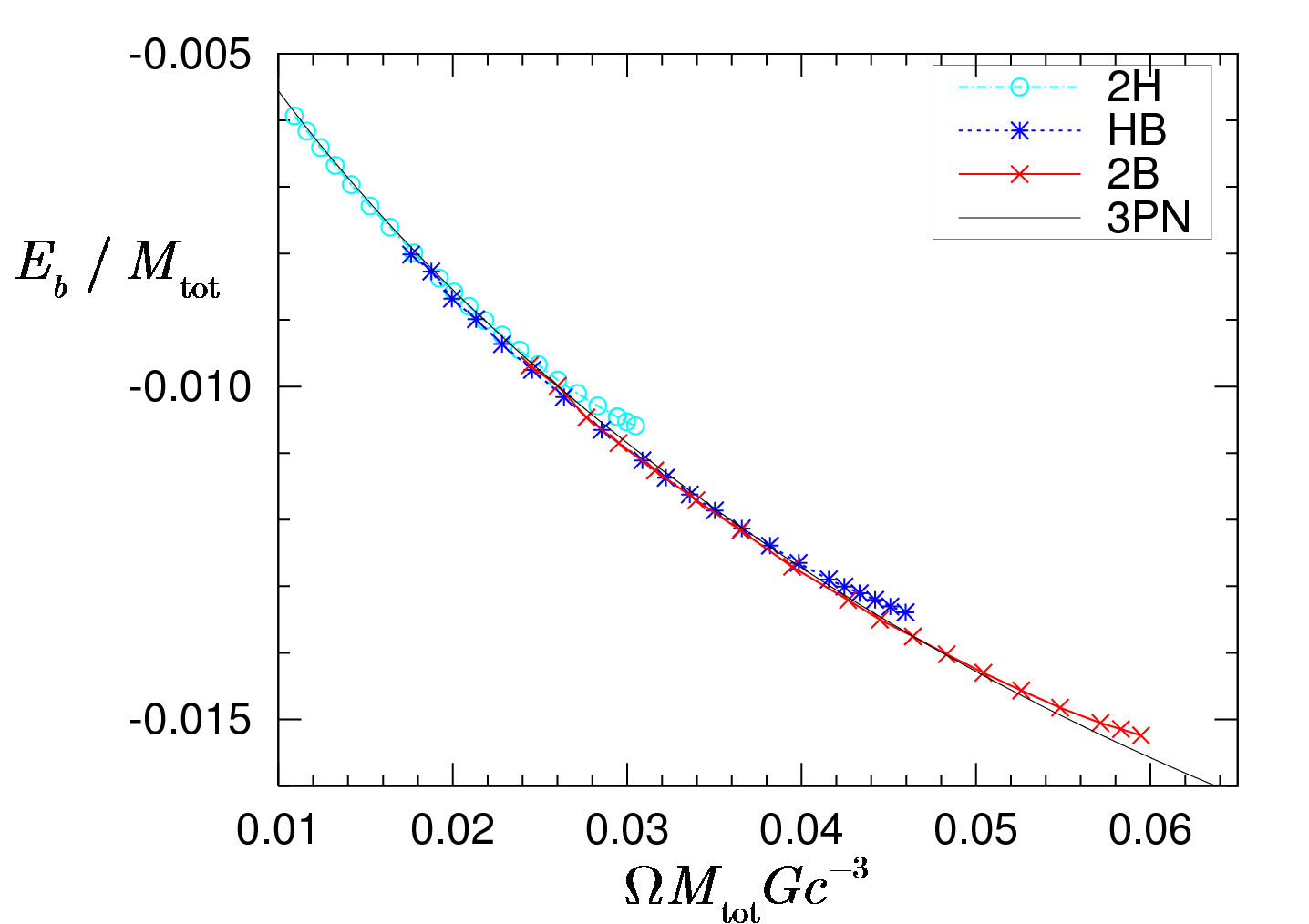} 
\end{center}
\caption[EOS for numerical evolution]{
\label{fig:qes1}\textbf{} Binding energy $E_{\text{b}} := M_{\text{ADM}} - M_{\rm{tot}}$ as a function of orbital frequency $\Omega$ for constant rest mass quasi-equilibrium sequences.   The gravitational mass of each neutron star in isolation is $M=M_{\rm{tot}}/2=1.35\,M_\odot$.
Departure from the point particle (3PN) curve 
occurs earlier for the stiffest  EOS (candidate 2H -- lowest compactness)  and later for the softest  EOS (candidate 2B -- highest compactness).}
\end{figure}

\subsection{Numerical evolution}

The Einstein equations are solved using the Baumgarte-Shapiro-Shibata-Nakamura
formulation~\cite{Shibata:1995we,bs99}. The conformal factor of the spatial
metric is evolved following \cite{Shibata:2007zm} and the resulting black
hole spacetime is handled using the moving puncture method~\cite{Campanelli:2005dd,bcckvm06}.
The lapse and shift are evolved using a dynamical gauge condition as in \cite{Shibata:2007zm},
while the Einstein equations coupled to the fluid equations are solved using the numerical scheme also described in  \cite{Shibata:2007zm,Readetal2009}.

During inspiral, the fluid evolution is essentially free of shocks, whence the cold parameterized EOS specified in
Sec.~\ref{sec:modeleos} is used in the simulations.  During merger, when shocks are developed, we include a hot EOS component with a thermal effective adiabatic
index $\Gamma_{\rm eff}$, as described in \cite{Shibata2003}.  Shock heating
in the merger can increase the thermal energy up to $\sim 20$--30\% of
the total energy \cite{Shibata:2005ss}.
The  merger and post-merger evolution is significantly affected by this hot component, since a  collapse to a black hole may be  delayed by the thermal pressure developed during and after merger.

Gravitational radiation is extracted both by spatially decomposing the
metric perturbation about flat spacetime in the wave-zone with spin-2
weighted spherical harmonics and by calculating the outgoing part of the
Weyl scalar $\Psi_4$.  For equal mass neutron stars, such as those
studied here, the quadrupole $(\ell=2,m=\pm2)$ mode is much larger
than any other mode, so we consider only this mode in this analysis.
The waveforms output from the simulations are the cross and plus
amplitudes $h_{+} c^2 D/GM_{\text{tot}}$ and $h_{\times} c^2
D/GM_{\text{tot}}$ of the quadrupole waveform, as would be measured at
large distance $D \gg GM_{\text{tot}}/c^2$ along the $z$ axis
perpendicular to the plane of the orbit, versus the retarded time
$t_{\text{ret}}$. Here $M_{\text{tot}}$ is the sum of the two neutron star
masses when they are far apart, $M_{\text{tot}}=2.7 M_\odot$.
%%The strain is sampled at discrete values evenly spaced in $t$, with a
%%sampling rate $\Delta t$ of between 0.006~ms (for 2B) and 0.031~ms
%%(for 2H) which depends on the time of simulation.
The early part of the extracted waveforms, for 
$t_{\text{ret}} \alt 0$, contains spurious
radiation which is discarded in the data analysis. 

Each simulation is typically performed for three grid resolutions.  In the
best-resolution case, the diameter of each neutron star is covered with 60
grid points. Convergence tests with different grid resolutions
indicate that, with the best grid resolution, the time duration in the
inspiral phase is underestimated by about 1 orbit. This is primarily
due to the fact that angular momentum is spuriously lost by numerical
dissipation. Thus, the inspiral gravitational waves include a phase
error, and as a result, the amplitude of the spectrum for the inspiral
phase is slightly underestimated. However, it is found that the waveforms
and resulting power spectrum for the late inspiral and merger phases, which
we are most interested in for the present work, depend weakly on the grid
resolution.

\section{Analysis of inspiral waveforms}

From the quadrupole waveform data, we construct the complex quantity 
\begin{equation}h = h_{+} - i h_{\times}
\end{equation}
The amplitude and phase of this quantity define the instantaneous amplitude
$|h|$ and phase $\phi = \arg h$ of the waveform. The
instantaneous frequency $f$ of the quadrupole waveform is then estimated by
\begin{equation}
f = \frac{1}{2 \pi} \frac{\Delta \phi}{\Delta t}
\end{equation}
The numerical data can be shifted in phase and time by adding a time shift
$\tau$ to the time series points and multiplying the complex $h$ by $e^{i
\phi}$ to shift the overall wave phase by $\phi$.

It is useful to define a reference time marking the end of the inspiral and
onset of merger. A natural choice for the end of the inspiral portion, 
considering the behaviour of a PP inspiral waveform, is the time of the peak
in the waveform amplitude $|h|$.
However, the amplitude of the numerical waveforms oscillates over the
course of an orbit. 
% (presumably due to orbital
%eccentricity, although extraction radius deformation could also
%contribute). 
A moving average of the waveform amplitude over $0.5$\,ms segments is used
to average this oscillation; the end of the inspiral is then defined as the
time at the end of the maximum amplitude interval. The resulting merger
time $t_{\text{M}}$ will be marked by solid vertical lines in the plots to
follow.

The numerical waveforms begin at different orbital frequencies. To align
them for comparison, they are each matched in the early inspiral region to
the same post-Newtonian point-particle (or PP) waveform.

\begin{figure}[!htb]

\begin{center}
\includegraphics[width=85mm]{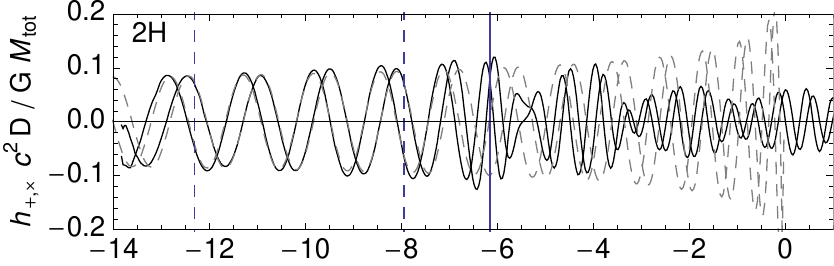} \\
\includegraphics[width=85mm]{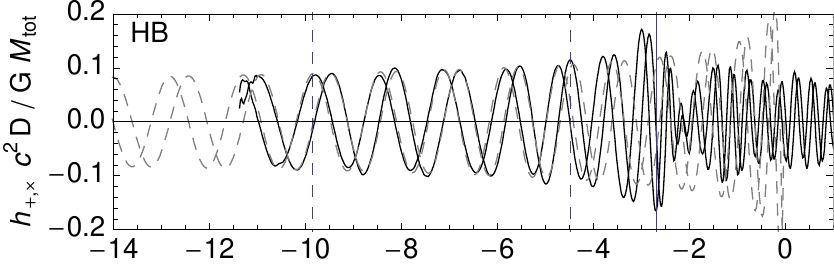} \\
\includegraphics[width=85mm]{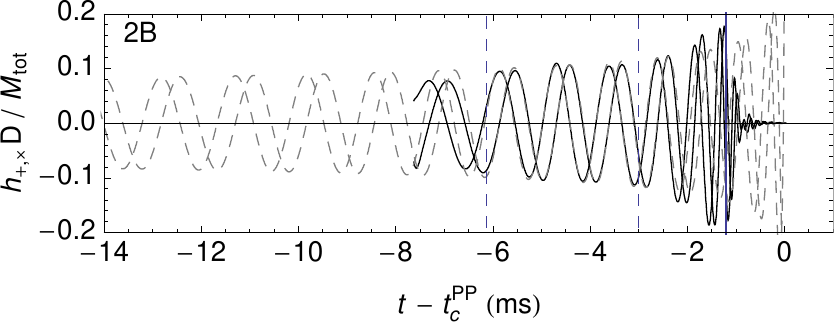}
\end{center}
\caption[Numerical waveforms aligned by PN match]{Solid lines show
numerical waveforms, scaled by
$c^2D/GM_{\text{tot}}$, and aligned in time and phase to the same
point-particle post-Newtonian inspiral (dashed line), using a method
described in~\cite{Readetal2009}. The two dashed vertical bars
indicate the portion of the waveform used for matching; the last vertical
bar indicates the end of inspiral time $t_{\text{M}}$ for the numerical
waveform.  The top
 %four 
two simulations, 2H
%, H, HB,   and B,
and HB,
show the start of
post-merger oscillations from a hypermassive neutron star remnant in the simulation.
2B shows quasinormal ringdown from a prompt collapse to a black hole
following merger.
Reproduced with permission from  \cite{Readetal2009} \copyright (2009) by the American Physical Society.\label{fig:matchalign}
}
\end{figure}

\subsection{Post-Newtonian point particle}

Point-particle inspiral is not well-defined in full general relativity (GR), and one is left
with the post-Newtonian point-particle (PP) approximation and fully general
relativistic black-hole numerical solutions as natural substitutes.
Fortunately, the Taylor T4 3.0/3.5 post-Newtonian specification, introduced in
\cite{Boyle:2007}, agrees closely with numerical binary black hole
waveforms for many cycles, up to and including the cycle before merger (see
also \cite{Gopakumar:2007}).  This empirical agreement allows us to adopt
the Taylor T4 waveform as an appropriate PP baseline waveform, compatible
with full GR until the last cycles. Quasi-equilibrium sequences [Fig. \ref{fig:qes1}] and time-frequency analysis [Fig. \ref{fig:timefreq}]
show that the binary
neutron star waveforms depart from this waveform $4$--$8$ cycles (200--560
$M_{\text{tot}}$) before the best-fit PP merger,  due to finite size effects.

The TaylorT4 waveform is constructed by numerically integrating to obtain a
gauge invariant post-Newtonian parameter  related to the orbital frequency
observed at infinity $\Omega$ \cite{Readetal2009}.
The orbital frequency evolution $\Omega(t)$ and orbital phase evolution  $\Phi(t)$ are computed to 3.5 post-Newtonian order following \cite{Boyle:2007}.
For the numerical integration, one needs to specify 
the constants of integration by fixing coalescence time
$t_{\text{c}}^{\text{PP}}$ and the orbital phase at
this time $\Phi(t_{\text{c}}^{\text{PP}})=\Phi_{\text{c}}^{\text{PP}}$. These two parameters uniquely specify the 3PN waveform for given particle masses.
The amplitude of
the ($l=2, m=\pm2$) quadrupole waveform is then calculated to 3.0 post-Newtonian order as described in 
\cite{Kidder:2008,Readetal2009}.

To match the numerical data to the PP inspiral waveform, 
the two parameters,
$t_{\text{c}}^{\text{PP}}$ and $\Phi_{\text{c}}^{\text{PP}}$ are 
varied and the best match is obtained \cite{Readetal2009}, to fix a relative time shift
and a relative phase shift. The masses of the point particles in the PP
waveform are fixed to be the same as the neutron stars in the numerical
simulations (the gravitational mass of isolated spherical neutron stars
with the same number of baryons) and so masses are not varied in finding
the best match.  With a goal of signal analysis, we choose the time and
phase shift by maximizing a correlation-based match between two waveforms.
More details on the matching technique may be found in \cite{Readetal2009}.
We note, however, that longer simulations are required to more
precisely fix a post-Newtonian match.

\subsection{Comparison of gravitational waveforms}

%
%\begin{figure}[!htb]
%\caption{ Time-amplitude behaviour, vertical line
%markings as previous
%figure.
%\label{fig:timeamp}}
%\begin{center}
%\begin{tabular}{cc}
%\includegraphics[width=60mm]{timeamp2B.pdf} \\
%\includegraphics[width=60mm]{timeamp5B.pdf} \\
%\includegraphics[width=60mm]{timeamp1B.pdf} \\
%\includegraphics[width=60mm]{timeamp4B.pdf} \\
%\includegraphics[width=60mm]{timeamp3B.pdf}
%\end{tabular}
%\end{center}
%\end{figure}

Unlike the case of matching binary black hole simulations to point particle
post-Newtonian\cite{Boyle:2007,Gopakumar:2007}, the binary neutron star simulations show departure
from point particle many cycles before the post-Newtonian merger time.
Fig.~\ref{fig:matchalign} shows the four waveforms shifted so the
best-match PP waveforms have the same $t_{\text{c}}^{\text{PP}}$ and
$\phi_{\text{c}}^{\text{PP}}$. As the stiffness of the EOS and thus
the radius of the neutron stars, increases, the end of inspiral for the
binary neutron stars is shifted away from the end of inspiral for post-Newtonian point
particle.

\begin{figure}[!htb]

\begin{center}
\begin{tabular}{cc}
\includegraphics[width=60mm]{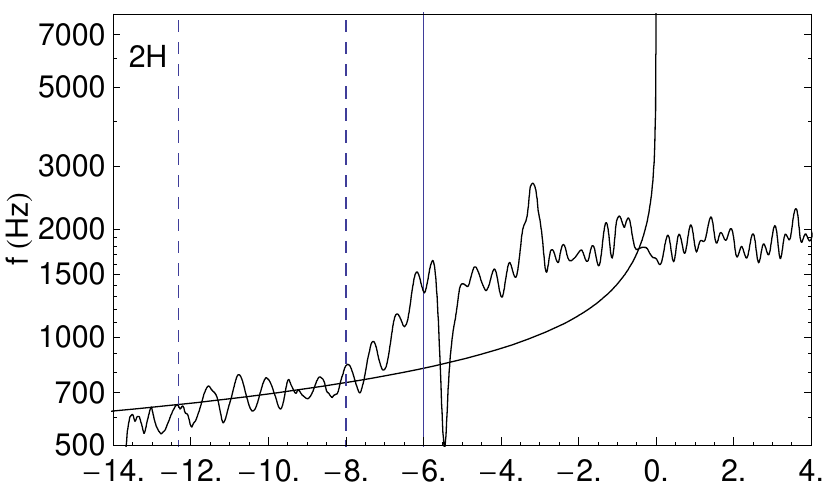} \\
\includegraphics[width=60mm]{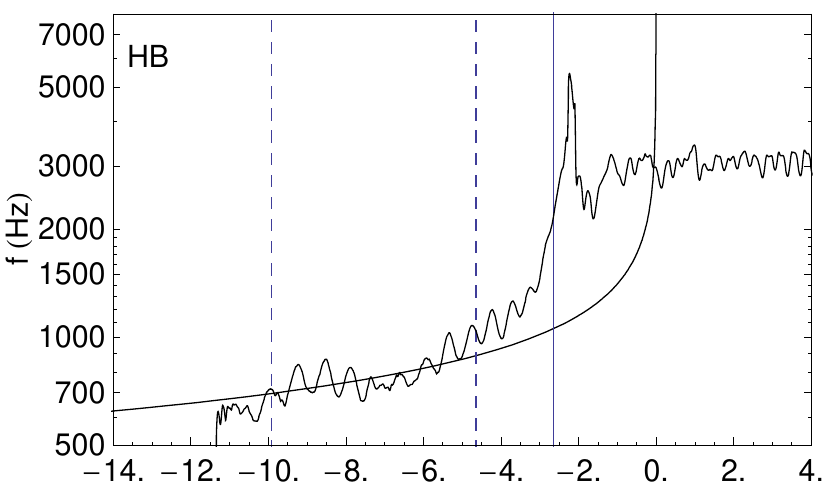} \\
\includegraphics[width=60mm]{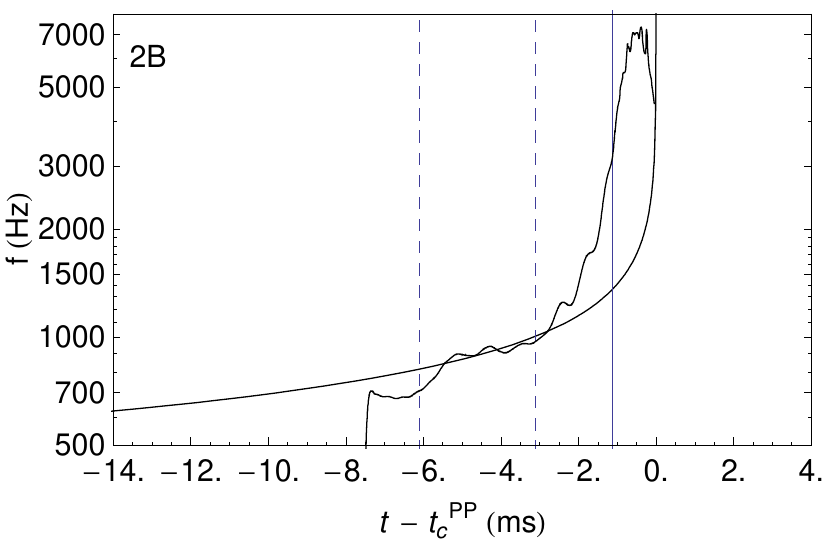}
\end{tabular}
\end{center}
\caption{Time-frequency behavior, vertical line
markings as previous figure. The departure from the point particle
time-frequency relations, shown using a long-dashed line, occurs between
$700$--$1000$\,Hz depending on the EOS.
In agreement with intuition and with Fig. \ref{fig:qes1}, the departure 
occurs earlier for the stiffest and later for the softest  EOS. Reproduced with permission from  \cite{Readetal2009} \copyright (2009) by the American Physical Society.\label{fig:timefreq}}
\end{figure}

This can also be seen by
plotting the instantaneous frequency of the numerical simulation waveform
with the same time shifts, as in Fig.~\ref{fig:timefreq}, which also
shows more clearly the difference in the post-merger oscillation
frequencies of the hypermassive remnants, when present. The larger neutron
star produced by the stiff EOS 2H has a lower oscillation frequency than
that from the medium EOS HB. 
%[[Are they all hypermassive? - check 2H max
%mass, it is super-stiff so maximum mass may be very large, check amount of
%matter thrown off realistic eos runs.]] 
The remnant forms with a bar-mode
oscillation stable over a longer period than the $\sim10$\,ms % [[refine]]
simulated. The signal from such a bar mode may be even stronger when the
full lifetime is included. Information that can be extracted
from the presence (or absence) and characteristics of a post-merger
oscillation signal from the hyper-massive remnant would complement the information present in the late
inspiral. Although  such information could probe
the equation of state at densities higher than those present in ordinary neutron stars, aspects of the physics neglected in
this study will likely come into play.
Such aspects may include finite temperature,  magnetic field, neutrino cooling and other effects that need to be accurately modeled, thus complicating the parameter extraction. As previously noted, our simulations included a hot component in the equation of state, which strongly affects the post merger behavior (and prompt versus delayed collapse to a black hole in particular), but inclusion of other effects is a subject for future study.

\subsection{Spectrum of gravitational waveforms}
Given $h_+$ or $h_\times$, one can construct the discrete Fourier
transforms (DFTs) $\tilde{h}_+$ or $\tilde{h}_\times$. Both polarizations
yield the same DFT amplitude spectrum $|\tilde{h}|$, with phase shifted by
$\pi/2$, if one neglects discretization, windowing, and numerical effects
(including eccentricity).  The amplitude spectrum $|\tilde{h}|$ is
independent of phase and time shifts of the waveform.
%[[[double check this]]

The stationary phase approximation is valid for the post-Newtonian waveform
up to frequencies of about 1500\,Hz (with $\lesssim 10\%$ error), so is used to
plot the amplitude of the point particle spectrum. In terms of an
instantaneous frequency
\begin{equation}f(t) = \frac{1}{2 \pi} \frac{d\phi}{dt},\end{equation}
the Fourier transform of the waveform has the amplitude
\begin{equation}
|\tilde{h}| \simeq A(f) \left(\frac{df}{dt}\right)^{-1/2} .
%\mathrm{exp}\left(i\left(2\pi
%f t_{\text{c}} - \phi(f) - \frac{\pi}{4}\right)\right).
\end{equation}
For binaries comprised of equal mass companions, the gravitational radiation
is dominated by quadrupole modes throughout the inspiral, so the analysis includes this mode only, as mentioned earlier.  The wave phase
$\phi$ is negligibly different from twice the orbital phase $2\Phi$ until
the onset of merger at very high frequencies so we can use the relation
\begin{equation}
\frac{df}{dt} \simeq \frac{1}{\pi}\frac{d\Omega}{dt};
\end{equation}
to write the amplitude of the Fourier transform entirely in
terms of the functions $d\Omega/dt$ and the amplitude
 $A(f) = |h|$
of the polarization waveforms.

%%\begin{equation}
%%\frac{df}{dt} \simeq \frac{1}{\pi}\frac{c^3}{G M} \frac{3}{2} x^{1/2}
%%\frac{dx}{dt};
%%\end{equation}
%%to write the amplitude of the Fourier transform entirely in
%%terms of the functions $dx/dt$ of Eq.~(\ref{eq:dxdt1}) and the amplitude
%%of the polarization waveforms $A(f) = |h|$ in
%%Eq.~(\ref{eq:h3pn}).

\begin{comment}
\begin{figure*} \label{fig:figdft}
\begin{tabular}{ll}
\includegraphics[width=70mm]{DFT2.pdf} 
%\includegraphics[width=70mm]{DFT5.pdf}\\ 
\includegraphics[width=70mm]{DFT1.pdf}\\
%\includegraphics[width=70mm]{DFT4.pdf}\\ 
\includegraphics[width=70mm]{DFT3.pdf}
\includegraphics[width=70mm]{InspiralDFTmerged.png} \\
\end{tabular}
\caption[DFT of full numerical waveforms]{DFT of full numerical waveforms,
at an effective distance $D_{\text{eff}}=100$\,Mpc, compared to
noise spectra for Advanced LIGO (labelled ``AdvLIGO'' for the standard
configuration and ``Broadband'' for the broad-band configuration) and the
Einstein Telescope (labelled ``ET'') shown by thick grey lines. The DFT of the
numerical waveforms turned off after the end of inspiral,
$t_{\text{M}}$, is shown by dot-dashed lines, the stationary-phase point particle is shown by a dashed line for
reference. The lower right figure shows a combined plot of
inspiral-truncated waveforms, smoothly joined on to best-match PP inspiral
time series before the DFT is taken. }
\end{figure*}
\end{comment}

\begin{figure*} \label{fig:figdft}
\begin{center}
\includegraphics[width=130mm]{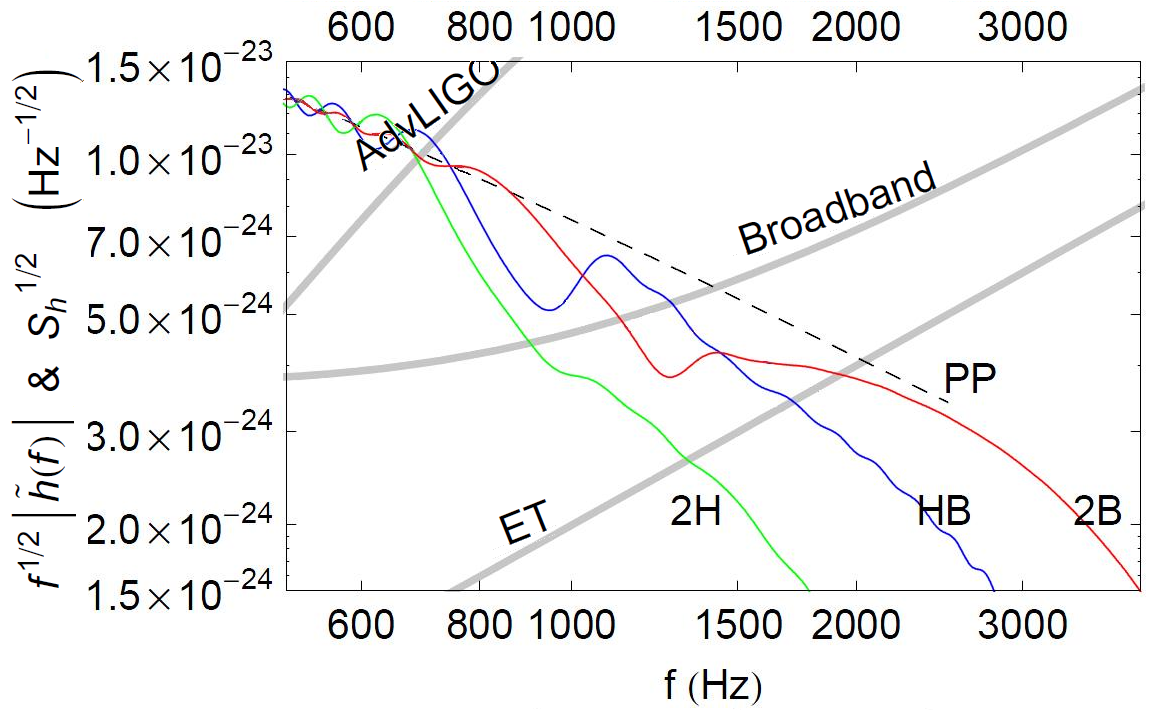}
\end{center}
\caption[DFT of full numerical waveforms]{Colored curves show  the DFT of  full numerical inspiral waveforms, truncated and smoothly joined  to best-match PP inspiral
time series,
at an effective distance $D_{\text{eff}}=100$\,Mpc. The stationary-phase point particle is shown by a dashed line for
reference. Thick grey lines represent the noise spectra for Advanced LIGO (labelled ``AdvLIGO'' for the standard
configuration and ``Broadband'' for the broad-band configuration) and the
Einstein Telescope (labelled ``ET''). Reproduced with permission from  \cite{Readetal2009} \copyright (2009) by the American Physical Society.}
\end{figure*}

The translation of emitted waveforms into the strain amplitude measured at
a detector involves transformations incorporating the effects of the
emitting binary's angle of inclination and sky location. 
%[[ For a quadrupole
%waveform, such as those considered here, the plus and cross polarizations
%at the detector are straightforwardly related somethiing something maybe]] 
These effects are absorbed into an effective distance $D_{\text{eff}}$,
which is equal to the actual distance $D$ for a binary with  optimal
orientation and sky location, and is greater than the actual distance for
a system that is not optimally oriented or located.
The detector will detect a single polarization of the waveform, some
combination of the plus and cross polarizations of the emitted waveform.
The polarizations extracted from the simulation can be used as two
estimates of the strain at the detector for a given numerically modelled
source, which give very close results and are subsequently averaged.

To compare with noise curves in the usual units, the quantity $f^{1/2}
|\tilde{h}(f)|$ is plotted, at a reference distance of
$D_{\text{eff}}=100$\,Mpc and rescaling from previously plotted numerical
output $h(t) c^2D/GM_{\text{tot}}$ using $M_{\text{tot}} = 2.7 M_\odot$.

The full spectra of models 2H and HB, seen in Fig.~7,
show peaks at post-merger oscillation frequencies. Waveforms of 2H and HB are truncated while the
post-merger oscillation is ongoing; if the simulations were allowed to
continue, these peaks would presumably grow further.  The simulation of the softest candidate 2B,
in contrast, promptly collapses to a black hole upon merger and has a short lived, and
relatively small-amplitude, quasinormal mode ringdown.

Note that time-frequency plots like the ones in Fig.~\ref{fig:timefreq}
showed that numerical waveforms follow the PP waveform at instantaneous
frequencies of up to $700$--$1000$\,Hz, depending on the EOS\@. The
disagreement in the spectra in Fig.~7 from the PP
stationary-phase approximation waveform at frequencies below this is
primarily due to the finite starting time of the numerical waveforms.  To
estimate spectra from the full inspiral, we construct hybrid waveforms.
The short-term numerical waveforms are  smoothly merged on to
long-inspiral PP-PN waveforms, using Hann windowing to smoothly turn on 
%\begin{equation}
%w(n) = \frac{1}{2}\left[1-\cos\left(\frac{\pi
%n}{N-1}\right)\right]
%\end{equation}
or turn off
%\begin{equation}
%w(n) = \frac{1}{2}\left[1+\cos\left(\frac{\pi n}{N-1}\right)\right]
%\end{equation}
a signal.
%over a range of $N$ points, $0\le n<N$ (or over a time $N\,\Delta t$).
To construct hybrid waveforms, these windows are used to turn on the
numerical waveform as the matched post-Newtonian is turned off, such that
the sum of the two window functions is 1 over the
matching range. Because, as mentioned earlier, the
dependence on the cold EOS is less straightforward during  the merger, due to unmodelled physics that become important at  that time, we focus
instead on the signal from the waveforms during the inspiral region only,
turning off the waveforms after the end of inspiral, $t_{\text{M}}$. 

%We see here t\uline{hat HB
%is estimated to depart from PP earlier than either H or B, rather than at
%the expected intermediate value}. This is ascribed to the higher
%eccentricity and lower accuracy of the match for the HB waveform.

\section{Distinguishability and detectability}
The question for neutron star astrophysics is whether these
differences in the gravitational waveform will be  measurable. We consider the
possibility of detecting EOS effects with Advanced LIGO style detectors
in varying configurations.

We use several detector configurations commonly considered for Advanced LIGO
include tunings optimized for 1.4~$M_\odot$ NS-NS inspiral detection
(``Standard''), for burst detection (``Broadband''), and for pulsars at
1150\,Hz (``Narrowband'').  The sensitivity is expressed in terms of the
one-sided strain-equivalent amplitude spectral density $S_h(f)$ (which has
units of $\mbox{Hz}^{-1/2}$) of the instrumental noise in Advanced LIGO\@.
We also consider a provisional noise curve for the Einstein Telescope
\cite{ETnoise}. We consider only a single detector of each type, rather
than a combination of detectors, for a preliminary estimate of
detectability.
 
Given two signals $h_1$ and $h_2$, and a
noise spectrum $S_h(f)$, we define the usual inner product \cite{CutlerFlanagan1994}:
\begin{equation}
\label{eq:sigmetr}
\langle h_1 | h_2 \rangle = 4 \mathop{\text{Re}}\nolimits\int_{0}^{\infty} \frac{
\tilde{h}_1(f) \tilde{h}_2^*(f) }{S_h(f)} df .
\end{equation}
This inner product yields a natural metric on a space of waveforms with
distance between waveforms weighted by the inverse of the noise. The detector output $s$ is filtered against an expected waveform $h$ using $\langle s | h \rangle$. Then
\begin{equation}
\varrho =\frac{ \langle s | h \rangle}{\sqrt{\langle h | h
\rangle}}
\end{equation}
is the optimal statistic to detect a waveform of known form $h$ in the
signal $s$.  If the detector output contains a particular waveform $h$
that is exactly matched by the template used, then the expectation value
of $\varrho$ is the expected signal-to-noise ratio or SNR of that waveform:
\begin{equation}
\overline{\varrho} = \sqrt{\langle h | h \rangle}.
\end{equation}

Given a signal $h_1$ to be used as a template, one
can ask whether a known \emph{departure} from this signal can be
measured.  Assuming the modified waveform $h_2$ is known, the expected
SNR of $h_2 - h_1$ is similarly
\begin{equation} \label{eq:rhodiffav}
\overline{\varrho}_{\text{diff}} = \sqrt{\langle h_2 - h_1 | h_2 -
h_1 \rangle}.
\end{equation}
We consider the two signals to be marginally distinguishable \cite{LindblomOwenBrown}
%\footnote{Compare
%discussion in \cite{LindblomOwenBrown} of
%indistinguishability.},
if the difference between the waveforms $h_1$ and $h_2$ has
$\overline{\varrho}_{\text{diff}} \geq
1$.

To analyze the measurability of finite size effects, we consider differences between hybrid
inspiral-only waveforms matched to point particle PN waveforms in the early
inspiral.  After the numerical waveforms have been matched to the same
post-Newtonian point-particle inspiral, the signals will be aligned in time
and phase.  We  then then compare the resulting waveforms to each other,
and to a PN-only waveform, using different Advanced LIGO noise spectra.
We report results in terms of the SNR measured by a single
detector at an effective distance $D_{\text{eff}}=100$\,Mpc. These results may be straightforwardly scaled to any distance in the wave zone, since $h$ and  $\overline{\varrho}_{\text{diff}}$ are inversely proportional to the distance $D$. 
%\begin{equation}
%\varrho_{\mathrm{diff},D} = \varrho_{\mathrm{diff},
%100\,\mathrm{Mpc}} \left( \frac{100\,\mathrm{Mpc}}{D} \right).
%\end{equation}

%\begin{table}[!htb]
\begin{table}[t]
\caption{$\overline{\varrho}_{\text{diff}}$ in standard (NS-NS detection optimized) noise $\times
\left( 100\,\mbox{Mpc} / D_{\text{eff}} \right)$ \label{tab:rdstd}}
\begin{center}
\begin{tabular}{c c c c c}\hline
Model 
&PP
&  2B 
%& B 
& HB 
%& H
 &2H\\
 \hline\hline
PP & 0 &0.32 
%&    0.45 
&  0.55 
%&  0.46
 &  0.69 \\
2B & & 0
%& 0.36 
& 0.48
% &  0.38
   &  0.63\\
%B  &  & 0& 0.21 &       0.12     &0.58\\
HB &  
%& 
& &0
%& 0.27
 & 0.60\\ 
%H & &  & & 0 & 0.58\\
%
2H &  
%& 
& & 
%& 0.27
 & 0\\ 
%H & &  & & 0 & 0.58\\
\hline
\end{tabular}
\end{center}
\end{table}

%\begin{table}[!htb]
\begin{table}[t]
\caption{$\overline{\varrho}_{\text{diff}}$ in broadband (burst-optimized) noise $\times \left(
100\,\mbox{Mpc} / D_{\text{eff}} \right)$ \label{tab:rdbb}}
\begin{center}

\begin{tabular}{c c c c c}\hline
Model 
&PP
&  2B 
%& B 
& HB 
%& H
 &2H\\
 \hline\hline
PP & 0 &1.86 
%&    0.45 
&  2.67 
%&  0.46
 &  2.89 \\
2B & & 0
%& 0.36 
& 2.32
% &  0.38
   &  2.54\\
%B  &  & 0& 0.21 &       0.12     &0.58\\
HB &  
%& 
& &0
%& 0.27
 & 2.37\\ 
%H & &  & & 0 & 0.58\\
%
2H &  
%& 
& & 
%& 0.27
 & 0\\ 
%H & &  & & 0 & 0.58\\
\hline
\end{tabular}

%\begin{tabular}{c  c c c c c}\hline
%Model  & 2B & B & HB & H&2H\\
%\hline\hline
%PP &1.86& 2.32& 2.67& 2.38& 2.89\\
%2B & 0&1.92& 2.32& 2.03& 2.54\\
%B  & & 0& 0.81& 0.80& 2.27\\
%HB  & & & 0& 1.28& 2.37\\
%H  & & & & 0&  2.35 \\
%\hline
%\end{tabular}
\end{center}
\end{table}
%\begin{table}[!htb]
\begin{table}[t]
\caption{$\overline{\varrho}_{\text{diff}}$ in narrowband $1150$\,Hz noise $\times \left(
100\,\mbox{Mpc} / D_{\text{eff}} \right)$ \label{tab:rdnb} }
\begin{center}

\begin{tabular}{c c c c c}\hline
Model 
&PP
&  2B 
%& B 
& HB 
%& H
 &2H\\
 \hline\hline
PP & 0 &0.91 
%&    0.45 
&  3.69 
%&  0.46
 &  2.12 \\
2B & & 0
%& 0.36 
& 2.92
% &  0.38
   &  1.45\\
%B  &  & 0& 0.21 &       0.12     &0.58\\
HB &  
%& 
& &0
%& 0.27
 & 2.25\\ 
%H & &  & & 0 & 0.58\\
%
2H &  
%& 
& & 
%& 0.27
 & 0\\ 
%H & &  & & 0 & 0.58\\
\hline
\end{tabular}

%\begin{tabular}{c c c c c c c} \hline
%Model &  2B & B & HB & H &2H\\ \hline\hline
%PP &  0.91& 2.75& 3.69& 2.65& 2.12\\
%2B & 0&1.92& 2.92& 1.82& 1.45\\
%B & & 0& 1.14& 0.22& 1.43\\
%HB &  & & 0& 1.34 & 2.25 \\
%H &  & & & 0&  1.42 \\
%\hline
%\end{tabular} 

\end{center}
\end{table} 

The difference between waveforms due to finite size effects is not
detectable in the NS-NS detection optimized configuration of Advanced LIGO
for $\sim 100$\,Mpc effective distances. 

However, in both narrowband and
broadband the differences can be significant, and waveforms are
distinguishable from each other and from the PP waveform. Note that this
implies that, even if the details of a NS-NS late inspiral signal are not
known, the difference between it and a point particle waveform should be
measurable.  The quantity $\overline{\varrho}_{\text{diff}}$ between the
observed waveform and a best fit point particle waveform, limited to
differences at high frequency, may be  useful  in itself to constrain
possible EOS independently of waveform details.

%Given an effective distance $D_eff$, the volume
%of space which is enclosed within this distance is [[blah]]. [[Something
%about multiple detectors.]]

\subsection{Parameter extraction}
Since the early part of the inspiral is well described by a point particle post-Newtonian approximation, we will assume that EOS effects on the waveform impact the late inspiral
only.  For simplicity, we assume that orbital parameters, such as
$M_{\text{tot}}$, mass ratio $\eta$, point particle  post-Newtonian
coalescence time $t_{\text{c}}^{\text{PP}}$, and phase shift
$\phi_{\text{c}}^{\text{PP}}$, are determined from the observations of the
earlier inspiral waveform, with
sufficient accuracy that their measurement uncertainty will not affect the
accuracy to which the late inspiral effects determine the EOS parameters.
These measurements would be made by a broad-band instrument, in which the
signal-to-noise ratio is expected to be high ($\sim 40$ at 100\,Mpc) and
measurement accuracy is expected to be good~\cite{CutlerFlanagan1994}.
Inaccuracies in these measurement could lead to biases in the measured
EOS\@.  This will be an important aspect to assess when
high-quality binary neutron star simulations with various masses become
abundant.

With a one-parameter family of waveforms sampled, we can estimate the
accuracy to which this parameter can be measured.
There are also other EOS-related parameters which are not considered. In
this first analysis we directly estimate the measurability of the EOS parameter $p_1$, 
ignoring variations of $\Gamma$ within the core. In the analysis of radius
measurability, variations of the internal structures are respectively neglected. 
Expanding coverage of the EOS parameter space is underway.
However, these initial parameter choices are expected to give the dominant
contributions to finite size effects of the waveform.
For example, the 1PN tidal effect estimates of
\cite{FlanaganHinderer2007,Hinderer2008} predict $\sim$10\% variation in
the tidal terms contributing to binding energy and luminosity from changing
internal structure---varying the apsidal constant---while keeping radius
fixed.

We estimate errors in parameter
estimation to first order in
$1/\overline{\varrho}$ or, equivalently, in $\delta\theta^A$, using the 
Fisher matrix $\Gamma_{AB} = \langle
\partial_A h | \partial_B h\rangle$ \cite{CutlerFlanagan1994}. Its inverse, $(\Gamma^{-1})^{AB}$,
yields
\begin{equation}
\overline{\delta \theta^A \delta \theta^B} = (\Gamma^{-1})^{AB}
\end{equation}
so that the expected error in a given parameter $\theta^A$ is
\begin{equation}
\overline{\left(\delta \theta^A \right)^2} = (\Gamma^{-1})^{AA}
\end{equation}
and the cross terms of the inverse Fisher matrix yield correlations between
different parameters.

With a few simulations of varying parameter value, we 
estimate  $\partial h/\partial {p_1}$ and $\partial h/\partial {R}$ from
two of the sampled waveforms, $h_1$ and $h_2$.  For a single parameter $\theta$
(which can be taken to be either $p_1$ or $R$), we have
\begin{equation}
\left. \frac{\partial h}{\partial {\theta} } \right|_{\theta=(\theta_{1}+\theta_{2})/2} 
\simeq \frac{ h_2 - h_1 }{\theta_{2} - \theta_{1}}
\end{equation}
where $h_1=h(\theta_1)$ and $h_2=h(\theta_{2})$,
and then, for our one-parameter family where we neglect correlations with other
parameters, we have to first order
\begin{equation} \label{eq:parameterrmserror}
\overline{\left(\delta \theta \right)^2} \simeq 
\frac{(\theta_{2} - \theta_{1})^2}
{\langle h_2 - h_1 | h_2 - h_1  \rangle}.
\end{equation}

Using adjacent pairs of models to estimate waveform dependence at an average
parameter value, we then find estimates of radius measurability as shown in
Table~\ref{tab:deltar} and $p_1$ measurability as shown in
Table~\ref{tab:deltap} for the burst-optimized noise and narrowband configurations.

%\begin{table}[!htb]
%
%
%%\begin{table}[t]
%%\caption{
%%Radius $R$ (km) measurability in broadband (burst-optimized) and narrowband
%%noise for a system at effective distance $100\,\mbox{Mpc}$.
%%The error $\delta R$ (km) is estimated using Eq.  \eqref{eq:parameterrmserror}
%%and scales with effective  distance  as
%%$ \delta R \times \left(
%%D_{\text{eff}}/ 100\,\mbox{Mpc}  \right)$
%%\label{tab:deltar}}
%%\begin{center}
%%\begin{tabular}{c c c c c c}\hline
%%Model & 2B & B& H               &2H\\
%%\hline\hline
%%2B & -- &0.63   & 1.28& 2.17\\
%%B  &            & --    & 1.74& 1.89\\
%%H  &            &                       & --    &  1.23\\
%%\hline
%%%\end{tabular}
%%\end{center}
%%\end{table}

\begin{table}[t]
\caption{
Measurability of radius $R$ (km)  with broadband (burst-optimized) and narrowband (1150~Hz) configurations of Advanced LIGO, for a binary system at effective distance $100\,\mbox{Mpc}$.
The values of $R$ here are midpoint values of the radii of candidates shown in Table \ref{tab:modprop}. The error $\delta R$ (km) is estimated using Eqs. 
\eqref{eq:rhodiffav},~\eqref{eq:parameterrmserror} (with $\overline{\varrho}_{\text{diff}}$ taken from Tables
\ref{tab:rdbb},~\ref{tab:rdnb})
and scales with effective  distance  as
$ \delta R \times \left(
D_{\text{eff}}/ 100\,\mbox{Mpc}  \right)$.\\
\label{tab:deltar}}
\begin{center}
\begin{tabular}{c c c}\hline
Configuration: 
%%& Standard
& Broadband& Narrowband (1150~Hz) \\
\hline\hline
$R=10.65$ 
%%& $\pm3.95$ 
& $\pm0.81$ &$\pm0.65$\\
$R=13.40$
%% & $\pm1.41$
 & $\pm1.52$ &$\pm1.60$\\
\hline
\end{tabular}
\end{center}
\end{table}

%\begin{table}[!htb]
\begin{table}[t]
\caption{ Measurability of pressure $p_1$ ($\mbox{dyn}\,\mbox{cm}^{-2}$)
at the fiducial density $\rho_1 = 10^{14.7}$\,g\,cm$^{-3}$,
with broadband (burst-optimized) and narrowband (1150~Hz) configurations of Advanced LIGO, for a binary system at effective distance $100\,\mbox{Mpc}$.
The values of $\mathrm{log}_{10}p_1$ here are midpoint values of $\mathrm{log}_{10}p_1$  from Table \ref{tab:modprop}. The error $\delta \mathrm{log}_{10}p_1$ is estimated using Eqs. 
\eqref{eq:rhodiffav},~\eqref{eq:parameterrmserror}
(with $\overline{\varrho}_{\text{diff}}$ taken from Tables
\ref{tab:rdbb},~\ref{tab:rdnb}) and scales with effective  distance  as
$ \delta \mathrm{log}_{10}p_1 \times \left(
D_{\text{eff}}/ 100\,\mbox{Mpc} \right)$. \\ \label{tab:deltap}}
\begin{center}
\begin{tabular}{c c c}\hline
Model & Broadband & Narrowband (1150~Hz) \\
\hline\hline
$ \mathrm{log}_{10}p_1=34.25$
& $\pm0.13$  &$\pm 0.10$\\
$ \mathrm{log}_{10}p_1=34.65$
& $\pm0.21$  &$\pm 0.22$\\
\hline
\end{tabular}
\end{center}
\end{table}

\subsection{Sources of error}

Many higher order but likely relevant 
effects have been neglected in this preliminary analysis.
Tidal effects may measurably influence the orbital evolution
before the start of the numerical simulations, as estimated in
\cite{FlanaganHinderer2007}, slowly enough not to be seen over the few
cycles of the waveform matched to PP in this analysis. In one sense this
analysis is a worst-case scenario, as it assumes exact PP behavior before
the numerical match.  Earlier drift away from point particle dynamics would
give larger differences between waveforms, and more sensitive radius
measurement, but poses a challenge by requiring accurate numerical
simulation over many cycles to verify EOS effects. Combining numerical
estimation with PN analyses like those of \cite{FlanaganHinderer2007,HindererLackeyRead2009},
and/or quasiequilibrium sequence information [Fig.~5] may clarify the transition
between effectively PP and tidally influenced regimes.

Some residual eccentricity from initial data
and finite numerical resolution is present in the waveforms themselves. The eccentricity error may be estimated by
comparing the plus polarization to the cross polarization shifted by
$\pi/2$ from the same numerical waveform. This results in
$\overline{\varrho}_{\text{diff}}$ of $\sim 0.3$ for HB and 2H, rather than
the expected quadrupole polarization cross-correlation of zero.  The 2B
waveform produces $\overline{\varrho}_{\text{diff}} \sim 0.05$ between
polarizations.
 
\begin{comment}
\sout{The value of
$\overline{\varrho}_{\text{diff}}$ for HB-B and H-HB should be half that of
H-B, but instead they are the same or greater---we are hitting the limit of
numerical and matching accuracy.  We can also estimate the validity of the
linear parameter dependence in the central models by comparing HB to
$\left( \mbox{H} +\mbox{B} \right)/2$. This results in
$\overline{\varrho}_{\text{diff}} \simeq  0.8$, another estimate of
systematic error.}  
\end{comment}

We have only a coarsely sampled family of waveforms; estimates of $\partial h/\partial \theta$ are limited by this. Further, the length of the inspirals (see discussion on PN matching in \cite{Readetal2009})
limits precision in choosing the best match time.  We can estimate these
effects on current results by varying the match region considered; this
changes $\overline{\varrho}_{\text{diff}}$  by up to $\sim 0.5$ at
100\,Mpc in the broadband detector. The resolution from existing numerical
simulations is comparable to the difference between parameters of some of the
 models.
The estimated uncertainty in each
$\overline{\varrho}_{\text{diff}}$ is smaller than (but comparable to) the $\overline{\varrho}_{\text{diff}}$ between candidates 2H, HB, 2B and PP.
 
Finally, we note that use  of a Fisher matrix estimate of parameter
measurement accuracy is fully valid only in high SNR limit of
$\overline{\varrho}_{\text{diff}} > 10$ \cite{vallisneri2007pe}, i.e.,
for distances $\lesssim 20$\,Mpc in the broadband detector.
The results do not take into account multiple detectors, nor multiple
observations, nor parameter correlations. We conclude that, although these are of course first estimates, they should
be better than order-of-magnitude. A full estimation of
EOS parameter measurability will require more detailed analysis, with a
larger set of inspiral simulations sampling a broader region of EOS parameter
space, with mass ratios departing from unity, and with more orbits
simulated before merger.

\section{Summary}

This is a first quantitative
estimate of the measurability of the EOS of cold matter above nuclear density with gravitational wave detectors. Neutron stars of the same mass but different EOS
and thus
different radii
will emit
%measurably
different gravitational waveforms during a late stage of binary inspiral. 
We calculated the signal
strength of this
difference in waveform using the sensitivity curves of commissioned and
proposed gravitational wave detectors, and find that there is a measurably
different
signal from the inspirals of binary
neutron stars with different EOS.  

Although the standard noise configuration of Advanced LIGO is not sensitive to
 differences in the waveform from finite size effects,
a
broadband (burst-optimized)  configuration or a 
narrowband detector configuration of Advanced LIGO can distinguish neutron star EOS from each other and from point particle inspiral, at an effective distance $\lesssim$ 100\,Mpc. In addition, the provisional standard noise curve of
the Einstein Telescope indicates the ability to resolve different EOS at
roughly double that distance. 

With broadband Advanced LIGO at frequencies between 500 and 1000\,Hz, our first estimates show that the radius can be determined to an accuracy of
$\delta R \sim 1\,\mbox{km} \times (100\,\mbox{Mpc}/D_{\text{eff}})$. Related first estimates show that such observations can constrain an EOS
parameter parameter $p_1$, the pressure 
at a rest mass density $\rho_1 =  10^{14.7}$\,g\,cm$^{-3}$,  with an accuracy of
$\delta p_1 \sim 5\times10^{33}\,\mbox{dyn}\,\mbox{cm}^{-2}\times (100\,\mbox{Mpc}/D_{\text{eff}})$.
These   estimates neglect
correlations between the parameters and other details of internal
structure, which are expected to give relatively small
tidal effect corrections.
 Although these results are preliminary, they strongly
motivate further work on gravitational wave constraints from binary neutron
star inspirals. The accuracy of the estimates will be improved  with  numerical simulation of more orbits in the late inspiral and a wider exploration of the EOS\ parameter space.

\section{Acknowledgments}
It is a pleasure to thank the organizers of the 13th Conference on Recent Developments in Gravity (NEB-\ XIII) at Thessaloniki,
for providing a very  stimulating environment for research and discussion during the meeting.  This work was supported in part by NSF grants PHY-0503366, PHY-0701817 and
PHY-0200852, by NASA grant ATP03-0001-0027, and by JSPS Grants-in-Aid for
Scientific Research(C) 20540275 and 19540263. CM thanks the Greek State 
Scholarships Foundation for support. Computation was done in part in the NAOJ and YITP systems.
%\bibliography{paper}

\end{document}